\documentclass[aps,prc,twocolumn,superscriptaddress,amsmath,showpacs,amssymb]{revtex4-1}
\usepackage{bm,physics}
\usepackage[dvipdfmx]{graphicx}

\usepackage{mathrsfs}

\newif\ifHIDEHIGHLIGNT
\HIDEHIGHLIGNTtrue 
\ifHIDEHIGHLIGNT
\usepackage{ulem,color}
\newcounter{nnn}

\else

\fi

\bmdefine{\ba}{a}
\bmdefine{\bb}{b}
\bmdefine{\bx}{x}
\bmdefine{\by}{y}
\bmdefine{\bz}{z}
\bmdefine{\bn}{n}
\bmdefine{\bp}{p}

\newcommand{\BM}{\begin{pmatrix}}
\newcommand{\EM}{\end{pmatrix}}

\renewcommand{\d}{\dagger}

\newcommand{\Lc}{\mathcal{L}}
\newcommand{\Mc}{\mathcal{M}}

\newcommand{\hphi}{\hat\varphi}
\newcommand{\hpsi}{\hat\psi}

\newcommand{\intx}{\int\! d^3x\;}
\newcommand{\intxd}{\int\! d^3x'\;}
\newcommand{\intxxd}{\int\! d^3x\,d^3x'\;}

\newcommand{\ex}{\mathrm{ex}}

\begin{document}
\title {
 Supersolidity  of    $\alpha$ cluster  structure  in $^{40}$C\lowercase{a}}

\author{S.~Ohkubo}
\affiliation{Research Center for Nuclear Physics,
Osaka University, Ibaraki,
Osaka 567-0047, Japan}

\date{\today}

\begin{abstract}
  $\alpha$ cluster structure in nuclei has been long understood based on the  geometrical  configuration picture.
  By using the  spatially localized Brink $\alpha$ cluster model 
    in the  generator coordinate method,  it is shown that  the    $\alpha$ cluster structure has the apparently opposing duality of crystallinity and condensation, a property of supersolids.
To study the condensation aspects of the   $\alpha$ cluster structure a field theoretical superfluid cluster model (SCM) is introduced, in which the order parameter of condensation is  incorporated by treating rigorously the Nambu-Goldstone mode due to spontaneous symmetry breaking of the global phase.
 The  $\alpha$ cluster structure of   $^{40}$Ca,  which has been understood in the crystallinity picture, is studied  by the SCM with ten $\alpha$ clusters.
 It is found that  the  $\alpha$ cluster structure of   $^{40}$Ca  is  reproduced by the SCM in addition to $^{12}$C reported in a previous paper, which gives support to  the duality  of the $\alpha$ cluster structure. 
 The emergence of the mysterious $0^+$ state at the lowest excitation energy near the $\alpha$ threshold energy  is understood to be a manifestation  of the Nambu-Goldstone zero mode, a soft mode,  due to the condensation aspect of the duality similar to the Hoyle state in $^{12}$C.
The  duality  of  $\alpha$ cluster structure with incompatible   crystallinity and   coherent wave nature due to condensation
    is the consequence of  the Pauli principle, which  causes clustering.
    \end{abstract}

\date{Received \today}
\maketitle
\par


\section{INTRODUCTION}  
\par
  A supersolid is a solid with superfluidity, and has been   sought   in
 recent decades in He II \cite{Andreev1969,Chester1970,Leggett1970,Matsuda1970,Boninsegni2012}. 
  Recently it was  created experimentally   in an optical lattice \cite{Leonard2017,Li2017,Tanzi2019,Bottcher2019,Chomaz2019,Tanzi2019A,Natale2019,Guo2019}. 
  The  observation of the Nambu-Goldstone mode  \cite{Nambu1960,Goldstone1960,Nambu1961} due to spontaneous symmetry breaking (SSB) of the global phase gives direct evidence of  supersolidity  for an optical lattice  \cite{Tanzi2019A,Natale2019,Guo2019}.
Very  recently  the existence of  supersolidity  of subatomic nature ---supersolidity of the three $\alpha$ cluster structure  in the nucleus $^{12}$C ---was  discussed \cite{Ohkubo2020}.

The $\alpha$ particle model, in which  the boson $\alpha$ particle  with spin 0 is considered as a constituent unit of the nucleus, was proposed as the first nuclear structure model in 1937 \cite{Wefelmeier1937,Wheeler1937,Wheeler1937B} but   
 criticized   \cite{Blatt1952}  in the advent of  the shell model \cite{Mayer1948,Haxel1949} and the collective model \cite{Bohr1952}. However, the successful  shell model and the collective model  \cite{Bohr1969A,Bohr1969B,Ring1980} also encountered   difficulty  explaining the emergence of  very low-lying  intruder  states in light  nuclei such as the mysterious  $0_2^+$ (6.05 MeV) state in $^{16}$O \cite{Arima1967,Marumori1968}. 
The    $\alpha$ cluster model based on the geometrical  crystallinity picture, in which the effect of the Pauli principle is taken into account, was revived and  has witnessed  tremendous success in recent decades  in explaining both shell-model like states and  $\alpha$ cluster states comprehensively, which are reviewed  for  light nuclei in Refs.   \cite{Suppl1972,Wildermuth1977,Suppl1980} and for  the medium-weight nuclei   in Ref. \cite{Suppl1998,Ohkubo1998}.

\par
The Brink cluster model based  on the  geometrical crystallinity picture using the  generator coordinate method (GCM) \cite{Brink1966}, the resonating group method (RGM), which is equivalent to the GCM \cite{Horiuchi1977},  and the orthogonality condition model (OCM) \cite{Saito1969} and the local potential model (LPM) \cite{Buck1975,Ohkubo1977,Michel1983,Ohkubo2016} ---both of which are  approximations of the RGM  and take into account of the Pauli principle by excluding the Pauli forbidden states in the RGM ---have been successful in understanding the structure of nuclei \cite{Suppl1972,Wildermuth1977,Suppl1980,Suppl1998,Ohkubo1998}. Examples are the  two $\alpha$ dumbbell structure of $^8$Be \cite{Horiuchi1970,Hiura1972}, the three $\alpha$ triangle structure of  $^{12}$C \cite{Uegaki1977,Uegaki1979,Kamimura1977}, and the $\alpha$+$^{16}$O structure in $^{20}$Ne \cite{Horiuchi1972,Hiura1972B,Fujiwara1980}.
   The unified understanding of  cluster structure in the  low energy region, and 
    prerainbows  and  nuclear rainbows in the scattering region, which are   confirmed for   systems such as $\alpha$+$^{16}$O and $\alpha$+$^{40}$Ca  \cite{Michel1998,Ohkubo2016}, supports   the geometrical crystallinity picture of the cluster structures.

  Very recently Ohkubo {\it et al.} \cite{Ohkubo2020} used a field theoretical superfluid  cluster model (SCM) for $^{12}$C  to  report that 
   the $\alpha$ cluster structure has a duality of crystallinity and condensation,  a property of supersolidity. 
According to this theory,   while the former is the view  from the particle nature of the cluster structure, the latter is the view from  the wave nature due to the coherence of the condensate cluster structure.
   It is  important to  clarify  whether this supersolidity of $\alpha$ cluster structure is inherent  only to the three $\alpha$ cluster structure of $^{12}$C or a general property of $\alpha$ cluster structure with geometrical crystallinity.
 $\alpha$ cluster structure was recently paid attention  from the viewpoint of  quantum phase transition \cite{Elhatisari2016,Ebran2020}.     

\par
In this paper, 
by using the   Brink $\alpha$ cluster model   it is shown generally  that     $\alpha$ cluster structure   has the duality of apparently  exclusive properties  of crystallinity and condensation, i.e., supersolidity.  The $\alpha$ cluster structure of $^{40}$Ca, which has been understood from the viewpoint of geometrical cluster structure, is studied  from the viewpoint of condensation, superfluidity, by using  a field theoretical superfluid $\alpha$ cluster model  which  treats rigorously spontaneous symmetry  breaking of the global phase due to condensation.  The mechanism of  why the mysterious $0^+$ state in $^{40}$Ca emerges as a collective state at very  low excitation energy, which has been a longstanding subject in the shell model and the collective model, is investigated  and is shown to arise as a member state of the  Nambu-Goldstone (NG) zero-mode due to global phase locking caused by the condensation aspect of the duality of  $\alpha$ clustering in $^{40}$Ca.
  
  \par
  The organization of this paper is as follows. In Sec. II  by using the Brink cluster model it is generally shown that  $\alpha$ cluster structure has a dual property of crystallinity and condensation. In Sec. III a field theoretical superfluid cluster model
    with the order parameter of condensation, in which   spontaneous symmetry breaking  of the global phase due to condensation is  rigorously treated,  is given.
  In Sec. IV $\alpha$ cluster structure  in  $^{40}$Ca is studied.    First,  historical attempts to understand the mysterious $0^+$ state  of $^{40}$Ca  in the shell model, collective model, and the $\alpha$ cluster model  are  briefly reviewed. It is summarized that by the observation of the $K=0^-$ band with the $\alpha$ cluster structure, which was  predicted from the study of  anomalous large angle scattering in $\alpha$ + $^{36}$Ar scattering, the geometrical $\alpha$ cluster model was confirmed.  
Second, from viewpoint of the duality of crystallinity and condensation, 
  the condensation aspect of the  $\alpha$ cluster of $^{40}$Ca is studied using  the superfluid cluster model.
   The mechanism of the emergence of a mysterious $0^+$ state is investigated. 
   Sec. V is devoted to discussion. A summary is given in Sec. VI.

\section{THE DUALITY OF THE  $\alpha$ CLUSTER STRUCTURE: CRYSTALLINITY AND CONDENSATION} 
\par
I show that $\alpha$ cluster structure with crystallinity  has condensate nature simultaneously by using the Brink $\alpha$ cluster model based on a  geometrical crystallinity picture.
 The $n$-$\alpha$ cluster model based on the  geometrical crystalline picture, such as the two $\alpha$ cluster model of $^8$Be and the  three $\alpha$ cluster model of $^{12}$C,  is given by the following Brink wave function \cite{Brink1966}:
  \begin{eqnarray}
& \Phi^{B}_{n\alpha}(\boldsymbol{R}_1, \cdots , \boldsymbol{R}_n)
=\frac{1}{\sqrt{(4n)!} }{\rm det} [\phi_{0s}(\boldsymbol{r}_1- \boldsymbol{R}_1) 
 \chi_{\tau_1,\sigma_1} \cdots \nonumber \\ 
&  \qquad\qquad\qquad\qquad \phi_{0s}(\boldsymbol{r}_{4n}-  \boldsymbol{R}_n)\chi_{\tau_{4n},\sigma_{4n}}],
 \label{Brinkwf}
 \end{eqnarray} 
  \noindent where   $\boldsymbol{R_i}$
  is a parameter that specifies the center  of the $i$th $\alpha$ cluster. 
 $\phi_{0s} (\boldsymbol{r} -  \boldsymbol{R})$ is a
0s harmonic  oscillator wave function with  a size parameter $b$  around a center  $\boldsymbol{R}$,
 \begin{align}
 \phi_{0s}(\boldsymbol{r} - \boldsymbol{R}) = \left( 
 \frac{1}{\pi b^2}\right)^{3/4} 
 \exp \left[  - \frac{ (\boldsymbol{r} - \boldsymbol{R})^2}{2b^2} \right], 
 \end{align}
 \noindent and  $\chi_{\tau,\sigma}$ is the spin-isospin wave function of a nucleon.
 Equation (\ref{Brinkwf}) is rewritten  as
  \begin{align}
 & \Phi^{B}_{n\alpha}(\boldsymbol{R}_1, \cdots , \boldsymbol{R}_n)  =\mathscr{A} \left[   \prod_{i=1}^n \exp \left\{   - 2 \frac{ (\boldsymbol{X}_i - \boldsymbol{R}_i )^2}{b^2}  \right\} \phi(\alpha_i)
  \right],
\label{Brinkwf2}
\end{align}
 \noindent where ${\boldsymbol X_i}$ is the center-of-mass coordinate  of the $i$th $\alpha$ cluster and   $\phi(\alpha_i$) represents the internal wave function of the $i$th $\alpha$ cluster. $\mathscr{A}$ is the antisymmerization operator.
 The generator coordinate wave function  $\Psi_{n\alpha}^{GCM}$ based on the geometrical configuration of the Brink wave function is  given by
  \begin{eqnarray}
 \Psi_{n\alpha}^{GCM} &=&\int d^3  \boldsymbol{R}_1 \cdots d^3 \boldsymbol{R}_n f(\boldsymbol{R}_1, \cdots , \boldsymbol{R}_n ) \nonumber \\
&    &\quad  \times\Phi^{B}_{n\alpha}(\boldsymbol{R}_1, \cdots ,  \boldsymbol{R}_n)\,.
 \label{GCMwf1}
 \label{GCMBrink}
\end{eqnarray}

\begin{figure*}[t] 
\includegraphics[width=17.2cm]{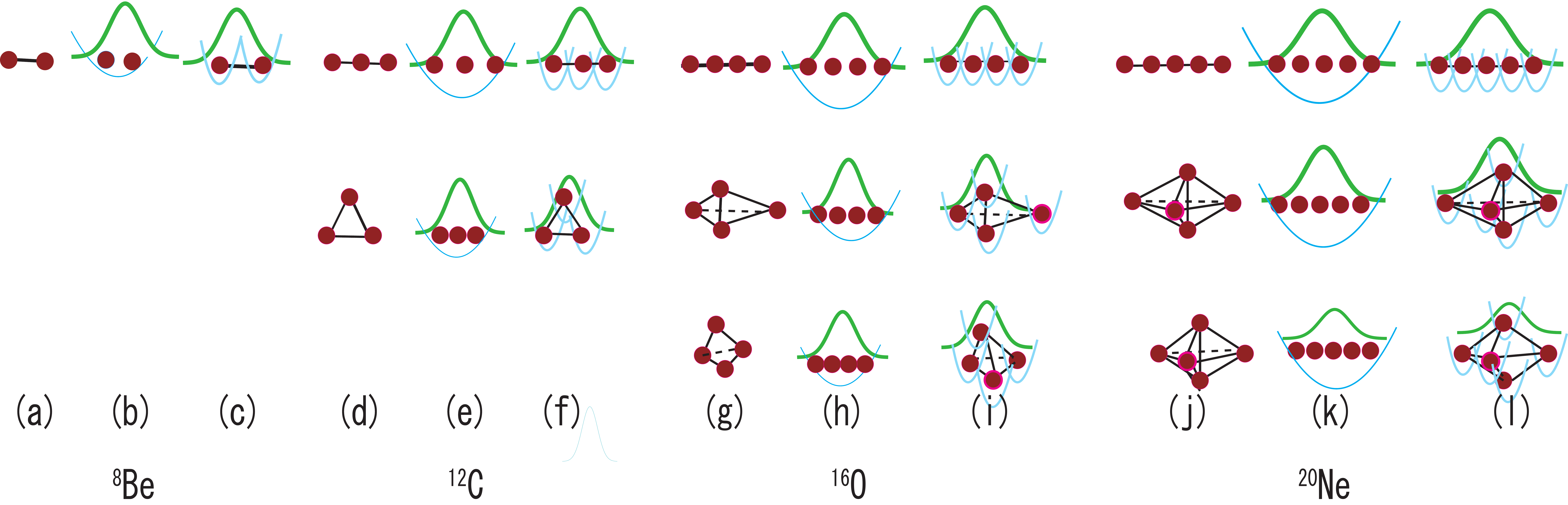}
 \protect\caption{
   Illustrative   figures  of  crystallinity,  condensation and supersolidity of the  $\alpha$ clusters  (filled circles)  proposed in $^{12}$C in Ref. \cite{Ohkubo2020} are displayed for the simplest system  $^8$Be discussed in the text and  extended to   $^{16}$O with four $\alpha$ clusters and $^{20}$Ne with   five  $\alpha$ clusters.   As the  excitation energy  increases   vertically,   the structure change occurs.  In  each nucleus   crystallinity,   condensation associated with a coherent wave,  and   supersolidity with both  crystallinity  and the coherent wave of the $\alpha$ clusters   are shown. The original Ikeda diagram based on the crystallinity picture corresponds to (a), (d), (g) and (j) in each nucleus. In (b), (e), (h) and (k) of each nucleus $\alpha$ clusters are sitting in the $0s$ state of the     harmonic oscillator potential with a  coherent   wave (broad curve).  In (c), (f), (i) and (l) of each nucleus, which illustrates supersolidity, the $\alpha$ clusters are sitting in the $0s$ state of the distinct harmonic oscillator potentials separated due to the Pauli repulsion associated with a  coherent   wave (broad curve).
 }
\label{fig1}
\end{figure*}

\par
I show that the  cluster model with crystallinity of Eq.~(\ref{GCMwf1})  has the property of condensation.
For the sake of simplicity I treat  hereafter
  the simplest two $\alpha$ cluster structure of  $^8$Be.
The generator  coordinates  $\boldsymbol{R}_1$ and  $\boldsymbol{R}_2$, which specify the position parameters of the two  $\alpha$ clusters, are rewritten   as follows by using $\boldsymbol{R}_G$ and $\boldsymbol{R}$, which are the center-of-mass  and  the relative vectors, respectively:
\begin{align} 
    \boldsymbol{R}_1 = \boldsymbol{R}_G +  \frac{1}{2} \boldsymbol{R}\,, \quad  
    \boldsymbol{R}_2 = \boldsymbol{R}_G -  \frac{1}{2} \boldsymbol{R}\,.
  \label{com}
\end{align}  
I take $\boldsymbol{R}_G$=0 to remove the spurious center-of-mass motion and use  the notation $\Phi^{B}_{2\alpha}(\boldsymbol{R})$  
 for $ \Phi^{B}_{2\alpha}(\boldsymbol{ \frac{1}{2}R},   \boldsymbol{ -\frac{1}{2}R})$.  Thus  Eq.~(\ref{GCMwf1}) is written as
 \begin{eqnarray} 
  \Psi_{2\alpha}^{GCM} & = & \int d^3   \boldsymbol{R} f(\boldsymbol {R}) 
\Phi^{B}_{2\alpha}(\boldsymbol{R})\,.
  \label{GCMwf8Be}
 \end{eqnarray} 
 \noindent  $\Psi_{2\alpha}^{GCM}$ is obtained by solving the Hill-Wheeler equation for  $f(\boldsymbol {R})$.
 I introduce ${g}(\boldsymbol{\mu})$, which is related to $f(\boldsymbol{R})$  by the Laplace transformation
 \begin{align} 
&    f(\boldsymbol {R}) = \int_{0}^{\infty} d\mu_x  \int_{0}^{\infty} d\mu_y \int_{0}^{\infty} d\mu_z 
 \exp\bigg[-(\mu_x {R_x}^2\nonumber \\
 &\qquad\qquad+\mu_y {R_y}^2+\mu_z {R_z}^2)\biggr] {g}(\boldsymbol{\mu}), 
\label{Laplace}
\end{align} 
\noindent where    $\boldsymbol{\mu}=(\mu_x, \mu_y, \mu_z)$.
Then Eq.(\ref{GCMwf8Be})  reads 
\begin{eqnarray} 
 \Psi_{2\alpha}^{GCM} & = &\int d^3   \boldsymbol{\mu} {g}(\boldsymbol {\mu}) \biggl[
 \int d^3  \boldsymbol{R} 
\exp \bigl\{-(\mu_x {R_x}^2+\mu_y {R_y}^2 \nonumber \\& & +\mu_z {R_z}^2)\bigr\}
\Phi^{B}_{2\alpha}(\boldsymbol{R})
\biggr].
  \label{GCMwf8Be2}
 \end{eqnarray} 
\noindent  By rewriting  the  term  $\left[\cdots\right]$ in the right-hand side  of Eq.
(\ref{GCMwf8Be2})
as ${\Phi}_{2\alpha}^{{PCM}}(\boldsymbol{\mu})$, defined by  
  \begin{align} 
& {\Phi}_{2\alpha}^{PCM}(\boldsymbol{\mu}) 
  \equiv  \int d^3  \boldsymbol{R}\exp \biggl[-(\mu_x {R_x}^2
  +\mu_y {R_y}^2\nonumber \\
  &\qquad\qquad\qquad+\mu_z {R_z}^2)\biggr]\Phi^{B}_{2\alpha}(\boldsymbol{R}),
  \label{NCMwf8Be2-1}
  \end{align} 
 \begin{align}
&  
\propto 
 \mathscr{A} \left[     \prod_{i=1}^{2}  \exp \left\{  - 2 \left(\frac{ X_{ix}^2}{B_{x}^2}+\frac{ X_{iy}^2}{B_{y}^2}+\frac{ X_{iz}^2}{B_{z}^2} \right) \right\}  \phi(\alpha_i)       \right],
\label{8BeNCM2}
\end{align}  
 \noindent  with $ B_{k} =\sqrt{b^2 + {\mu_{k}}^{-1} }   \;  (k=x, y, z)$, 
 Eq. (\ref{GCMwf8Be2})   reads 
  \begin{eqnarray}
  \Psi_{2\alpha}^{GCM} & = &\int d^3   \boldsymbol{\mu} {g}(\boldsymbol{\mu})  {\Phi}_{2\alpha}^{{PCM}}(\boldsymbol{\mu}).
 \label{GCMNCMwf8Be2-1}
 \end{eqnarray}
 The ${\Phi}_{2\alpha}^{{PCM}}$ in 
 Eq. (\ref{8BeNCM2}), which is { called  a nonlocalized cluster model (NCM) or a THSR cluster  wave
 function in  Refs. \cite{Tohsaki2001,Zhou2020}} {and  suggested to be a pseudo-condensate model (PCM) in Ref. \cite{Ohkubo2020}},  shows that  $\alpha$ clusters are  sitting in the $0s$ orbit
 of the trapping  harmonic oscillator potential with an oscillator parameter $\boldsymbol{B}=(B_x, B_y, B_z)$ and represents the condensed aspect of the  $\alpha$ clusters.
  This trapping harmonic oscillator potential corresponds to the trapping harmonic oscillator  potential in the superfluid 
  cluster model discussed in the next section where the condensation of the trapped $\alpha$  clusters is rigorously treated by introducing the order parameter due to SSB of the vacuum.
 While  in Eq.(\ref{GCMwf8Be}) the GCM wave function is expressed  based on the geometrical picture using the Brink function $\Phi^{B}_{2\alpha}(\boldsymbol{R})$ as a base function, in  Eq.(\ref{GCMNCMwf8Be2-1})   the same GCM wave function is expressed   using the  wave function ${\Phi}_{2\alpha}^{{PCM}}(\boldsymbol {\mu})$ as a base function.
 From Eqs.(\ref{GCMwf8Be}) and (\ref{GCMNCMwf8Be2-1}), it is found that
the  Brink $\alpha$ cluster model  in the  generator coordinate method  has both the crystallinity and the condensate 
nature simultaneously.

  \par
  The above discussion for the simplest two $\alpha$ cluster system 
 can be    generalized to the $n$-$\alpha$ cluster system.  
  The Laplace transformation relation    is   generalized to  
  \begin{eqnarray}   
    f(\boldsymbol{R}_1, \cdots , \boldsymbol{R}_n) & = &
\int_{0}^{\infty} 
   d\boldsymbol{\mu}
\exp\biggl[- \sum_{i=1}^{n} (\mu_x R_{ix}^2+\mu_y R_{iy}^2 \nonumber \\
&& +\mu_z R_{iz}^2)\biggr] {g}(\boldsymbol{\mu}). 
\label{Laplace2}
\end{eqnarray} 
Equation (\ref{NCMwf8Be2-1})    generalized to $n$-$\alpha$ clusters is given by 
\begin{align} 	
& \Phi_{n\alpha}^{PCM}(\boldsymbol{\mu})   = \int d^3  \boldsymbol{R}_1 \cdots d^3 \boldsymbol{R}_n  \exp \biggl[  -\sum_{i=1}^{n}  (\mu_{x} R_{ix}^2 \nonumber \\
& \qquad\qquad +\mu_{y} R _{iy}^2 + \mu_{z} R_{iz}^2)    \biggr] 
\Phi^{B}_{n\alpha}(\boldsymbol{R}_1, \cdots ,  \boldsymbol{R}_n),   
\label{NCM1}
\end{align} 
 \begin{align}
& \propto  \mathscr{A} \left[     \prod_{i=1}^{n}  \exp \left\{  - 2 \left(\frac{ X_{ix}^2}{B_{x}^2}+\frac{ X_{iy}^2}{B_{y}^2}+\frac{ X_{iz}^2}{B_{z}^2} \right) \right\} 
 \phi(\alpha_i)  
     \right].
\label{NCM2}
\end{align}
Similarly to  Eq.(\ref{GCMNCMwf8Be2-1}), one gets
\begin{align}
&  \Psi_{n\alpha}^{GCM}=\int d^3   \boldsymbol{\mu} {g}(\boldsymbol{\mu})  {\Phi}_{n\alpha}^{{PCM}}(\boldsymbol{\mu}).
 \label{GCMNCMwf-Nalpha}
  \end{align} 
   
 Thus from   Eqs. (\ref{GCMwf1}) and (\ref{GCMNCMwf-Nalpha})
it is found   generally  that  the  $n$-$\alpha$ cluster wave function
  in the  geometrical cluster model picture has the property of condensation.  
This   shows generally  that the GCM $n$-$\alpha$ cluster  wave function   has the duality of crystallinity
and condensation
  independently of the Hamiltonian used.

       Illustrative pictures based on  the above geometrical structure and the condensate structure  of the   $\alpha$ clusters in $^8$Be,  $^{12}$C, $^{16}$O, and $^{20}$Ne are displayed          in Fig.~1.   
 The pictures     (a), (d), (g), and (j) correspond to the Ikeda diagram  \cite{Ikeda1968,Horiuchi1972} based on the crystallinity.
      The pictures (b), (e), (h), and (k)  represent  the wave aspect of the $\alpha$ cluster structure due to the condensation.
     The two exclusive pictures, the duality of crystallinity and a condensate coherent wave,   can be      unified in the  pictures displayed  in (c), (f), (i), and (l) of each nucleus in Fig.~1  where the $\alpha$ clusters   sitting in the $0s$ state of  the distinct potentials due to the Pauli  repulsion   between the $\alpha$ clusters \cite{Tamagaki1968} form a coherent wave.

The de Broglie wavelength of each   $0s$ state $\alpha$ cluster with  very low energy    is far longer than the geometrical distance between the $\alpha$ clusters. In other word, the phases of the waves are locked to form a  coherent wave function, i.e.,  superfluidity (condensation) of the system.   This 
      is general and  independent of the geometrical configuration and  number of the  $\alpha$ clusters involved, $n$. Therefore  in principle, whatever the geometrical configuration is ---triangle ($n$=3) structure of $^{12}$C,  tetrahedron ($n$=4) structure of $^{16}$O \cite{Dennison1954,Bijker2014,Halcrow2017,Halcrow2019,Halcrow2020}, trigonal bipyramid ($n=$5) structure of $^{20}$Ne  \cite{Bouten1962,Brink1968,Brink1970,Nemoto1975,Bijker2021},  linear chain $n$-$\alpha$ cluster  ($n$=2, 3, 4,   $\cdots$), etc.---the geometrical $\alpha$ cluster structures have the potential  to form a coherent  wave function (superfluidity). Whether the   state  is superfluid  depends on  the superfluid density   $\rho_s$, which   encapsulates  the structure and degree of  clustering.
   The  previous  study  of $^{12}$C \cite{Ohkubo2020} finds that the superfluid  ground state is stable with a condensation rate  5\%,  giving  energy levels similar to the GCM, RGM  and experiment.

\section{ FIELD THEORETICAL SUPERFLUID CLUSTER MODEL FOR CONDENSATION OF  $\alpha$ CLUSTERS
 } 

The traditional cluster models  involve no order parameter that characterizes condensation.  A theory with no order parameter is unable to  conclude whether a system under investigation is  condensate or not.
 In  Eqs. (\ref{GCMNCMwf-Nalpha}) and (\ref{GCMwf1})  no order parameter to characterize the condensation is involved. 
  In  Eq. (\ref{GCMwf1}) the parameter  $\boldsymbol{R}$ is the order parameter to characterize the geometrical clustering. In fact, $\boldsymbol{R}=0$ corresponds to the shell model with no clustering and $\boldsymbol{R}\ne0$ represents the degree of geometrical clustering.  On the other hand, in Eq. (\ref{GCMNCMwf-Nalpha}) the parameter 
  $\boldsymbol{B}$   is not {self-evidently} the order parameter of condensation because   global phase locking caused by spontaneous symmetry breaking  due to condensation is not  {explicitly} involved.  
In fact,    $\boldsymbol{B}=0$ has no physical meaning  and 
   $\boldsymbol{B}\ne0$  does not necessarily  mean  condensation {in Eq.  (\ref{GCMNCMwf-Nalpha})}.
To conclude  whether the $\alpha$ cluster structure has Bose-Einstein condensate  (BEC) nature, it is necessary to use a theory in which  the  order parameter to characterize condensation is implemented.

I  briefly present  the formulation of the field theoretical superfluid  cluster model developed in \cite{Nakamura2016,Nakamura2018,Katsuragi2018} to study BEC of  $\alpha$ clusters in the  Hoyle state and the excited states above it in $^{12}$C.
The  model Hamiltonian for a bosonic field $\hpsi(x)$
$(x=(\bx,t))$ representing
the $\alpha$ cluster is given as follows:
\begin{align}
&\hat{H}=\intx \hpsi^\d(x) \left(-\frac{\nabla^2}{2m}+
V_\ex(\bx)- \mu \right) \hpsi(x) 
\notag\\
&\,\,
+\frac12 \intxxd \hpsi^\d(x)
\hpsi^\d(x') U(|\bx-\bx'|) \hpsi(x') \hpsi(x) \,.
\label{Hamiltonian}
\end{align}
Here, the potential $V_\ex$ is a mean field potential introduced phenomenologically 
to trap the $\alpha$ clusters inside the nucleus, and is taken to have a harmonic
oscillator form,
$
 V_\ex(r)= m \Omega^2 r^2/2\,.
$
 $U (|\bm x -\bm x'|)$  is the  residual $\alpha$--$\alpha$ interaction.
I set $\hbar=c=1$ throughout this paper.

\par
When BEC of $\alpha$ clusters occurs, i.e.,
the global phase symmetry of $\hpsi$ is spontaneously broken, I decompose $\hpsi$
as $\hpsi(x)=\xi(r)+\hphi(x)$, where the $c$-number $\xi(r)=\bra{0} \hat\psi(x)
\ket{0}$ is an order parameter and is assumed to be real and isotropic.
To obtain the excitation spectrum, one needs to solve
three coupled equations, which are the Gross--Pitaevskii (GP) equation, Bogoliubov-de Gennes (BdG) equation, and the zero-mode equation \cite{Nakamura2016, Katsuragi2018}.
The GP equation  determines the order parameter
   by
\begin{equation}\label{eq:GP}
\left\{ -\frac{\nabla^2}{2m}+V_\ex(r) -\mu + V_H(r)
\right\} \xi(r) = 0 \,,
\end{equation}
where
$
    V_H(r) = \intxd U(|\bx-\bx'|)\xi^2(r')\,.
$  The order parameter $\xi$ is normalized
with the  condensed particle number $N_0$ as
\begin{align}
\intx |\xi(r)|^2 = N_0\,.
\end{align}
The BdG equation  describes the collective oscillations on the condensate
 by
\begin{align}
  \intxd
  \left(\begin{array}{cc}
        \Lc & \Mc \\
        -\Mc^* & -\Lc^*
      \end{array}\right)
  \left(\begin{array}{c}
    u_{\bn} \\
    v_{\bn}
   \end{array}\right)
  = \omega_\bn
    \left(\begin{array}{c}
        u_{\bn} \\
        v_{\bn}
\end{array}  \right),
\end{align}
where
\begin{eqnarray}
  \Mc(\bx, \bx')
   &= &U(|\bx-\bx'|) \xi(r) \xi(r'),\, \nonumber\\
  \Lc(\bx, \bx')
   &=& \delta(\bx-\bx')
     \left\{ -\frac{\nabla^2}{2m}+V_\ex(r) -\mu + V_H(r) \right\}  \nonumber \\
     &&+ \Mc(\bx, \bx')\,. 
\end{eqnarray}
The index $\bm{n}=(n,\, \ell,\, m)$ stands for the main, azimuthal and
magnetic quantum numbers. The eigenvalue $\omega_\bn$ is the excitation energy
of the Bogoliubov mode. For isotropic $\xi$, the BdG eigenfunctions can be taken to have separable forms,
\begin{eqnarray}
u_{\bm{n}}(\bm{x}) &= &\mathcal{U}_{n\ell}(r) Y_{\ell m}(\theta, \phi), \,  \nonumber\\
 v_{\bm{n}}(\bm{x}) & = &\mathcal{V}_{n\ell}(r) Y_{\ell m}(\theta, \phi).
\label{BdGsolution}
\end{eqnarray}
One necessarily has an eigenfunction belonging
to zero eigenvalue, explicitly $(\xi(r), -\xi(r))^t$, and its adjoint function
$(\eta(r),\eta(r))^t$ is obtained as
\begin{equation}
\eta(r)= \frac{\partial}{\partial N_0}\xi(r)\,.
\end{equation}
The field operator is expanded as
\begin{align}
\hphi(x)&=-i{\hat Q}(t)\xi(r)+{\hat P}(t)\eta(r) 
+\sum_{\bn} \left\{{\hat a}u_\bn(\bx)
+{\hat a}^\dagger v^\ast_\bn(\bx) \right\}\,,
\end{align}
with the commutation relations $[{\hat Q}\,,\,{\hat P}]=i$ and
$[{\hat a}_{\bn}\,,\,{\hat a}^\dagger_{\bn'}]=
\delta_{\bn \bn'}$\,. The operator ${\hat a}_\bn$ is an annihilation operator
of the Bogoliubov mode, and the pair of canonical operators ${\hat Q}$ and ${\hat P}$
originate from the SSB of the global phase and are called the NG or zero-mode operators.

The treatment of the zero-mode operators is a chief feature of the present approach.
The naive choice of the unperturbed bilinear Hamiltonian with respect
to ${\hat Q}$ and ${\hat P}$ fails due to their large quantum fluctuations.
Instead,  all the terms consisting only of ${\hat Q}$ and ${\hat P}$
in the total Hamiltonian are gathered to construct the unperturbed nonlinear Hamiltonian
of ${\hat Q}$ and ${\hat P}$, denoted by $\hat H_u^{QP}$\, with
\begin{eqnarray} 
\hat H_u^{QP} &=&- \left(\delta\mu + 2C_{2002} + 2C_{1111} \right) \hat P+
\frac{I-4C_{1102}}{2}\hat P^2 \nonumber \\ 
 &&+ 2C_{2011}\hat Q\hat P\hat Q  
+ 2C_{1102}\hat P^3 
+ \frac{1}{2}C_{2020}\hat Q^4  \nonumber \\ 
&& -2C_{2011}
\hat Q^2  
+ C_{2002}\hat Q\hat P^2\hat Q + \frac{1}{2} C_{0202}\hat P^4\,,
\label{eq:HuQP}
\end{eqnarray}
where 
\begin{eqnarray}
C_{iji'j'} &= &
\int\! d^3x d^3x'\, U(|\bx-\bx'|) 
 \{\xi(\bx)\}^i
\{\eta(\bx)\}^j \nonumber \\
&& \times \{\xi(\bx')\}^{i'}\{\eta(\bx')\}^{j'}
\,,
\label{eq:Cijij}
\end{eqnarray}
and $\delta \mu $ is a counter term that the 
criterion $\bra0\hpsi\ket0=\xi\,$ determines. 
I set up the eigenequation
for $\hat{H}_u^{QP}$, called
the zero--mode equation,
\begin{align} \label{eq:HuQPeigen}
\hat H_u^{QP} \ket{\Psi_\nu} = E_\nu \ket{\Psi_\nu}\qquad
(\nu=0,1,\cdots)\,.
\end{align}
This equation is similar to a one-dimensional 
Schr\"odinger equation for a bound problem.

\par
The total unperturbed Hamiltonian ${\hat H}_u$ is ${\hat H}_u=\hat H_u^{QP}
+\sum_{\bn} \omega_\bn {\hat a}_\bn^\dagger{\hat a}_\bn$.
The ground state energy is set to zero, $E_0=0$.
The states that I consider are $\ket{\Psi_\nu}\ket{0}_{\rm ex}$ with
energy $E_\nu$, called the zero-mode state, and
$\ket{\Psi_0}{\hat a}^\dagger_\bn
\ket{0}_{\rm ex}$ with energy $\omega_\bn$, called the BdG state, 
where ${\hat a}_\bn \ket{0}_{\rm ex}=0$.

 \section{ $\alpha$ CLUSTER STRUCTURE IN   $^{40}$C\lowercase{a}}    
 
 In order to discuss the macroscopic concept of  condensation in  nuclei,  it seems    suitable to study a nucleus which involves   many $\alpha$ clusters.  I take    $^{40}$Ca with ten $\alpha$ clusters.
   In this section I apply the  field theoretical superfluid  cluster model to $\alpha$ cluster structure study of  $^{40}$Ca  with the mysterious  $0^+$ state at 3.35 MeV.

 First, I review briefly how 
 the mysterious $0^+$ state in the doubly magic nucleus has been
   understood from the viewpoint of mean field theory, i.e., the  shell model and the collective model, in the recent decades  and how the $\alpha$ cluster model based on the crystallinity  picture has explained the mysterious $0^+$ state.
Second,  I study whether the $\alpha$ cluster states explained in the geometrical configuration picture can be understood in the viewpoint of condensation  of the duality by using the SCM. 
 
 \subsection{ THE MYSTERIOUS   $0^+$ STATE AND GEOMETRICAL    $\alpha$ CLUSTER STRUCTURE OF     $^{40}$C\lowercase{a}} 

 \par
The emergence of the $0_2^+$ at the very low excitation energy 3.35 MeV of the doubly shell-closed  magic nucleus $^{40}$Ca  as well as  the  $0_2^+$ state at 6.05 MeV  in $^{16}$O  had been mysterious from the viewpoint of the shell model \cite{Brown1966,Gerace1967,Gerace1969,Arima1967}.
 Brown and Green \cite{Brown1966} pointed out the importance of   deformed four-particle four-hole (4p-4h) excitations in lowering the excitation energy of the  $0_2^+$ state in $^{16}$O.   Gerace and Green \cite{Gerace1967} showed that the same situation occurs for the   $0_2^+$ state in   $^{40}$Ca.  Gerace and Green \cite{Gerace1969} showed that 8p-8h excitation is important in understanding the third $0_3^+$ state at 5. 21 MeV.  
  In the shell model calculations in which  the $^{32}$S core is assumed, Sakakura {\it et al.} \cite{Sakakura1976} argued that the   $0_2^+$  and  $0_3^+$ states are dominated by the 4p-4h excitations  and   the $0_4^+$  state at 7.30 MeV is dominated by the 8p-8h excitation.

\par
These studies show  that  the vacuum ground state of $^{40}$Ca is not   a simple 0p-0h  spherical shell model state but involves a  non-negligible  amount of  4p-4h correlations. To understand the excited structure is to reveal  the correlations, the predisposition,  of the vacuum ground state. 
 From this viewpoint Marumori and Suzuki \cite{Marumori1968} suggested  to understand  the mechanism of the emergence of the mysterious $0^+$ state as a collective state   by defining a vacuum with correlations of the 4p-4h mode. Following  this idea, the  four-particle and  four-hole mode-mode coupling was investigated in  $^{16}$O \cite{Takada1974} and $^{40}$Ca \cite{Hasegawa1976}.

\par
 The nuclear energy density functional (EDF) approach has been applied to describe the ground state properties and the collective excitations including clustering, especially the microscopic analysis of the formation and evolution of the cluster structure  from the vacuum ground state. The authors of Ref. \cite{Ebran2014} consider that cluster structures can be a transitional phase between the quantum liquid phase and the crystal phase.  It is very interesting to know  whether  the mysterious $0^+$ states  in   $^{40}$Ca as well as in $^{16}$O are reproduced in  the EDF and how its mechanism  of low excitation energy is  understood from the viewpoint of the  mean field. However, the structure of the excited energy levels in   $^{40}$Ca  as well as   $^{16}$O have not been reported yet.

\par
Also {\it ab initio} approaches, such as fermionic molecular dynamics (FMD) \cite{Chernykh2007} used for  $^{12}$C and  antisymmetric molecular dynamics (AMD) \cite{Taniguchi2014} used for  $^{42}$Ca,  have not been applied to explain the mysterious $0^+$ state   of $^{40}$Ca.  In  $^{16}$O  {\it ab initio} calculations have been  unable to explain the very low excitation  energy of the mysterious $0^+$ state   providing an  excitation energy  13.3 MeV in Ref. \cite{Furutachi2008} and 19.8 MeV in Ref. \cite{Wloch2005}, which are two or three times larger than the experimental value, 6.05 MeV. 

\par
From the viewpoint of $\alpha$ cluster structure, 
 Ogawa, Suzuki and Ikeda \cite{Ogawa1977} investigated the structure of $^{40}$Ca using  the  $\alpha$+$^{36}$Ar  cluster model, in which
  no  $K=0^-$  band appears, which is  a parity-doublet partner of the $K=0^+$ band
 built on the mysterious $0_2^+$  state, was obtained. Since this situation
  looked very  different from  $^{16}$O where  the well-developed $\alpha$  cluster  $K=0^-$ band, which is a parity doublet partner 
of the $K=0^+$  band    built on the mysterious $0_2^+$ (6.05 MeV) state  \cite{Suzuki1976A,Suzuki1976B},  Fujiwara {\it et al.}  \cite{Fujiwara1980} discussed   that the   $K=0^+$ band   in $^{40}$Ca has rather strong shell model aspects than the  $\alpha$ cluster structure.

\par
On the other hand, from the viewpoint of unification of cluster structure  in the  bound and quasibound states 
and    backward angle anomaly (BAA) or anomalous large angle scattering (ALAS) in $\alpha$+$^{36}$Ar scattering,  Ohkubo and Umehara \cite{Ohkubo1988} showed  that the  $2^+$ (3.90 MeV), $4^+$ (5.28 MeV), and $6^+$ (6.93 MeV) states built on the mysterious $0_2^+$ state form a  $K=0^+$ band with the $\alpha$+$^{36}$Ar  cluster structure  and  predicted the existence of a parity-doublet partner  $K=0^-$ band with the well-developed $\alpha$ cluster structure at slightly above the $\alpha$ threshold energy.
 The observation  of  the predicted  $\alpha$ cluster $K=0^-$ band   by Yamaya {\it et al.} \cite{Yamaya1993,Yamaya1994,Yamaya1998} in an $\alpha$ transfer reaction experiment  showed  that  the $K=0^-$ band and  the $K=0^+$ band      have the $\alpha$ cluster structure.
   The  $\alpha$ spectroscopic  factor, $S^2_\alpha$=0.30,  extracted from  
    $\alpha$ transfer reactions \cite{Yamaya1993,Yamaya1994,Yamaya1998} shows that the ground state has a  significant $\alpha$ cluster correlation.
    The  $\alpha$ cluster structure of $^{40}$Ca was further  confirmed theoretically 
by the semi-microscopic $\alpha$ cluster model calculations using the orthogonality condition model   by Sakuda and Ohkubo \cite{Sakuda1994,Sakuda1998}, in which not only the $\alpha$ cluster model space but also the shell model space are incorporated. 
In the OCM calculations   not only the $\alpha$ cluster states but also the shell-model like states  in $^{40}$Ca are reproduced   in the $\alpha$+$^{36}$Ar cluster model.  

\par
  Thus  the mysterious $0^+$ state  of  $^{40}$Ca was  found to  emerge 
   from the   ground state with the predisposition of $\alpha$ clustering.
 The finding   that the vacuum ground state involves  $\alpha$ cluster correlations
  is  consistent with the shell model studies in Refs. \cite{Brown1966,Gerace1967,Gerace1969,Arima1967} and the collective model viewpoint in Refs. \cite{Marumori1968,Hasegawa1976}, which
suggests   that the ground state involves   multiparticle-multihole, dominantly 4p-4h, shell model   components.  
   The geometrical $\alpha$ cluster model has been also  successful in  describing  well the $\alpha$ cluster structure in the $^{40}$Ca - $^{44}$Ti region \cite{Ohkubo1998,Michel1998,Sakuda1998}.

\par
Recently Manton   \cite{Manton2020}  reported  that the energy levels of $^{40}$Ca  can be classified  as the vibration and rotation of the  ten $  \alpha$ clusters using  a Skyrme  model. 
Microscopic ten $\alpha$ cluster model  calculations using the RGM and the GCM as well as the semi-microscopic OCM  
may be desired, however, such ten-body calculations are far beyond  the  power of the modern supercomputers.  From the microscopic cluster model point of view the $\alpha$+$^{36}$Ar cluster  model is an approximation of the ten $\alpha$ cluster model with
 $\boldsymbol{R}_1= \boldsymbol{R}_2= \cdots  =\boldsymbol{R}_9$ in Eq.(\ref{GCMwf1}), as illustrated as in Fig.~2.  

\par
Since the crystallinity picture of the $\alpha$ cluster structure in $^{40}$Ca has been confirmed theoretically and experimentally,  
the problem is  to reveal the origin and the collective nature of the mysterious $0^+$ state as well as the excited $\alpha$ cluster states from the  condensation viewpoint of the duality.

\subsection{SUPERFLUID CLUSTER MODEL STUDY OF  $^{40}$Ca}  
Because  the $\alpha$  cluster structure invokes  the duality of  geometrical structure and condensation  as discussed  in Sec. II, it is expected  that the $\alpha$ cluster states in $^{40}$Ca can  be also understood   from the condensation viewpoint using the   SCM with  the order parameter of condensation.
In a previous paper \cite{Ohkubo2020}  the SCM was applied to understand  the duality of the $\alpha$ cluster structure  of $^{12}$C, for which    the $\alpha$ cluster condensation of the Hoyle state  had been  thoroughly investigated theoretically and experimentally.
In contrast to the computational  difficulties of the  ten $\alpha$ cluster GCM calculations,
  the SCM calculation is tractable  for many $\alpha$ cluster systems.
In fact, the SCM has been successfully  applied to study  the BEC  of $\alpha$ clusters at   high excitation energies in many nuclei, $^{12}$C, $^{16}$O, $^{20}$Ne, 
 etc., and in $^{48}$Cr and $^{52}$Fe with thirteen $\alpha$ clusters \cite{Katsuragi2018}.

\begin{figure}[t]
\begin{center}
\includegraphics[width=8.6cm]{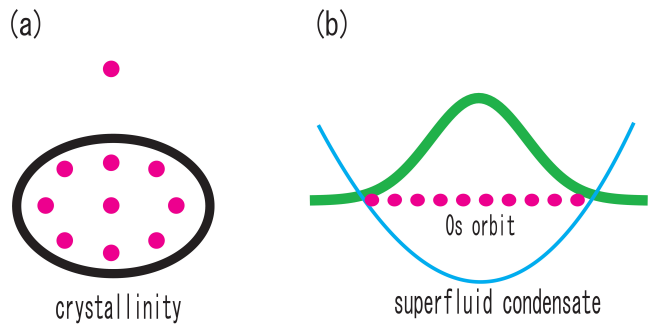}
\end{center}
\protect\caption{ (Color online) Illustrative   pictures of ten $\alpha$ clusters (filled red circles) in $^{40}$Ca.  (a) The crystallinity picture of the  $\alpha$ cluster structure of $^{40}$Ca with the $\alpha$+$^{36}$Ar  geometrical  configuration.  (b) The condensation picture of the  $\alpha$ cluster structure of $^{40}$Ca  in the superfluid  cluster model where the ten  $\alpha$ clusters associated a  coherent wave (broad  curve) are trapped in the $0s$ orbit of the confining  potential.
} 
\label{fig2}

\end{figure}

 \begin{figure}[!thb]
\begin{center}
\includegraphics[width=8.6cm]{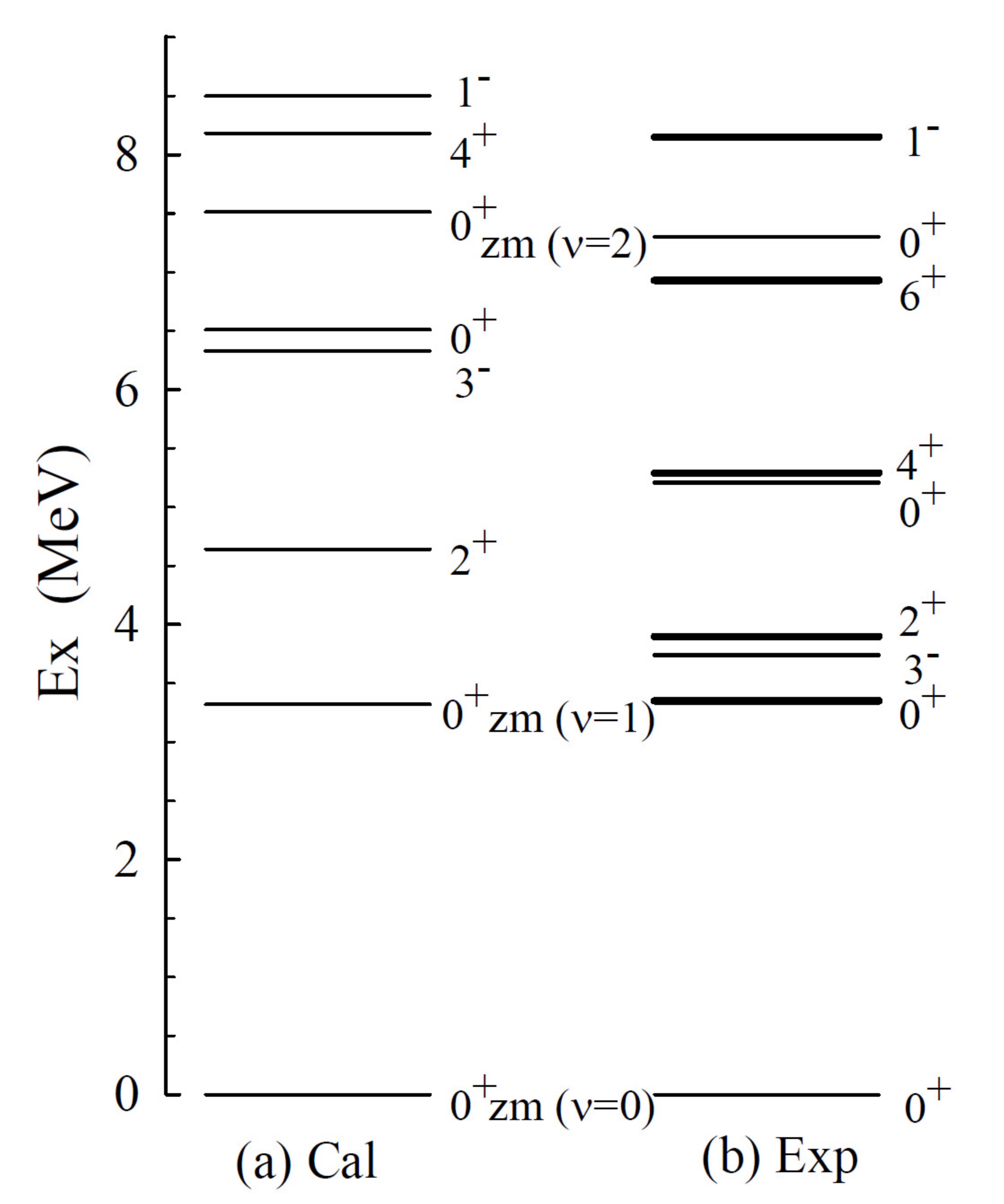}
\end{center}
\protect\caption {  (a) The energy levels  of $^{40}$Ca calculated  with the  superfluid   cluster model  with the condensation rate  6\%. The zero-mode states are indicated by zm and others are BdG states. (b) The  experimental low-lying energy levels of $^{40}$Ca \cite{Yamaya1994,Yamaya1998,ENSDF}. 
 The member states of the experimental  $K=0^+$ and $K=0^-$ bands  with the $\alpha$ cluster structure are indicated by the thick solid lines.  
 }
\label{fig:SCMEnergyLevel}

\end{figure}

As in Refs.~\cite{Ohkubo2020,Nakamura2016,Nakamura2018,Katsuragi2018,Takahashi2020}, I take the
  Ali--Bodmer type two-range Gaussian   nuclear potential $U (|\bm x -\bm x'|)$
 $ = V_r e^{-\mu_r^2 |\bm x -\bm x'|^2} - V_a e^{-\mu_a^2 |\bm x -\bm x'|^2}\,$, 
 with  $V_r$ and $V_a$ being  the     strength  parameters of the short-range  repulsive potential due to the Pauli principle  and the long-range attractive  potential, respectively \cite{Ali1966}.
 The chemical potential is fixed by the specification of the superfluid particle number $N_0$.
I assume the condensation rate to be  6\%, $N_0=0.06N$. 
The ground state  is identified  as the vacuum $\ket{\Psi_0}\ket{0}_{\rm ex}$.  The
range parameters $\mu_a$  and $\mu_r$ are fixed to the values $0.475$  and  $0.7$ ${\rm fm}^{-1}$, respectively, 
determined in Ref.   \cite{Ali1966}  to reproduce  $\alpha$+$\alpha$ scattering.
  The two potential parameters, $\Omega$, which controls the size of the system, and $ V_r$, which prevents  collapse of the condensate, are determined to be
 $\Omega=2.97 $ MeV$/ \hbar$ and $ V_r$=591 MeV. These
 reproduce the experimental root mean square (rms) radius 3.43  fm of the ground state, $\ket{\Psi_0}\ket{0}_{\rm ex}$ and   the   energy  level of the $0_2^+$  state  identified as the first excited   zero-mode  state $\ket{\Psi_1}\ket{0}_{\rm ex}$.
 
 \par
 In Fig.~3, the calculated energy levels   are compared with the experimental data.
The calculation locates  the $K=0^+$ band states in correspondence to  the experimental band build on the mysterious $0_2^+$ state. 
 The moment of inertia of the calculated band is smaller than the experimental one and the $6^+$ state appears at 12.75 MeV.  However,  it is to be noted  that the  $\alpha$ cluster band  emerges at  very low excitation energy from the spherical vacuum.   The  $0_2^+$ state is a state of the Nambu-Goldstone zero mode, a soft mode collective state. This soft mode nature  explains naturally   why the mysterious collective  $0^+$ state emerges at such a low excitation energy, although it is mysteriously low for a  4p-4h state in the shell model to emerge   from  the spherical vacuum ground state of $^{40}$Ca.  If  the system is infinitely large, it would  appear at  zero excitation energy.  The   finite low  excitation energy is the consequence of  the finite size of the nucleus and the Pauli principle.  The excited states $2^+$ and $4^+$  of the $K=0^+$ band are the BdG states built on the NG mode state. 
 
 The calculated  $0_3^+$ state, which is a BdG mode state, corresponds to the experimental   $0_3^+$ state at 5.21 MeV above the $\alpha$ threshold energy, which is considered to be an 8p-8h   state in the deformed  model \cite{Gerace1969} and a 4p-4h   dominant state in the $^{32}$S core shell model calculations \cite{Sakakura1976}.  The calculated  $0_4^+$  state, which is a second member state of the NG zero mode, corresponds to the experimental $0_4^+$ state at 7.30 MeV, which is interpreted to be  a   4p-4h and  8p-8h dominant state  in the shell model calculation \cite{Sakakura1976} and a  2p-2h dominant  state in the deformed  model \cite{Gerace1969}. As for the  negative state, a collective  $3^-$ state appears in accordance with experiment although the calculated energy is slightly  high. The $1^-$ state also appears to be in good  correspondence with the experimental energy level, which is considered to be a band head state of the parity doublet $K=0^-$ band.   It is to be noted  that, although  no geometrical configuration of the ten $\alpha$ clusters are   assumed, the important $\alpha$ cluster states are  obtained  in good accordance with experiment  by the SCM calculation based on the picture of Fig.~\ref{fig2}.

 \begin{figure}[t]
\includegraphics[width=8.6cm]{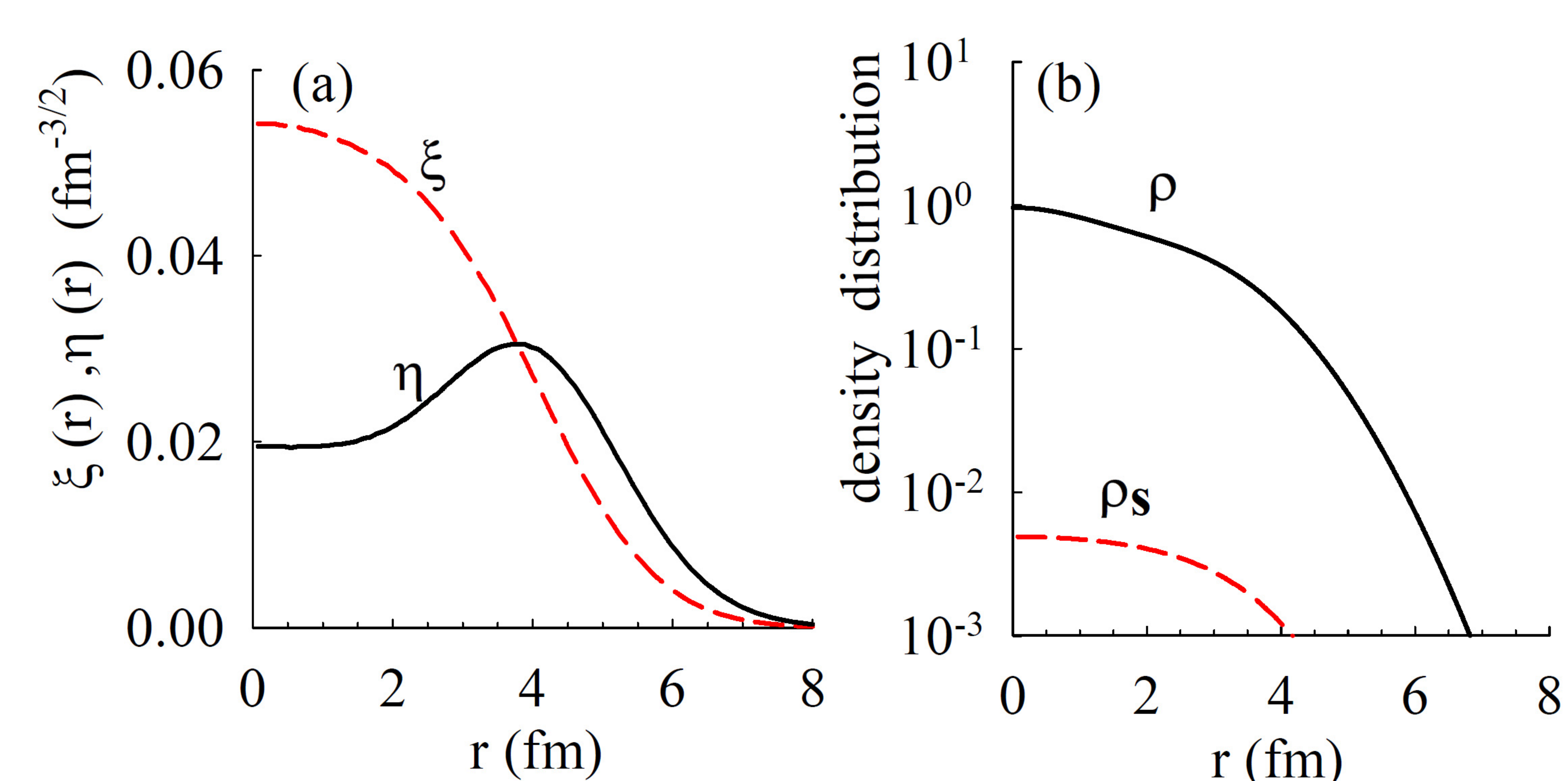}
\protect\caption{(Color online)  (a) The calculated eigenfunction  (order parameter)  $\xi (r)$ (dashed line) and its adjoint eigenfunction $\eta(r)$ (solid line) for the ground state of $^{40}$Ca.  (b)   The  calculated  superfluid density distribution   $\rho_s$
  in the SCM  (dashed line) and the matter density distribution  $\rho$  in the OCM of Refs. \cite{Sakuda1994} (solid line) for the ground state.
}
\label{fig:orderparameter}

\end{figure}

In Fig.~4(a) the calculated eigenfunction $\xi(r)$ (the order parameter) and   its adjoint eigenfunction $\eta(r)$ are displayed. The number fluctuation of the superfluid $\alpha$ clusters in the ground state, $\eta$, is large near
the surface region and decreases toward the inner and outer regions. In Fig. 4(b) 
$\rho_s(r)=|\xi(r)|^2/N_0$  represents
the calculated superfluid density distribution of  the  $\alpha$ clusters and $\rho(r)$ is the nuclear matter density distribution  calculated in the OCM cluster model of Refs. \cite{Sakuda1994}. $\rho_s$ is largest in the
center of the nucleus and gradually decreases toward the surface region. The non-superfluid normal
density,
 $\rho_n$ $\equiv$ $\rho$ -- $\rho_s$, is much smaller than $\rho$. However, it is this small superfluid
density component that causes the $0_2^+$ state at such low excitation energy as an NG zero-mode  state.
This may evoke that the Cooper pairs, a small fraction of nucleons near the Fermi surface, cause the superfluidity of  nuclei  in the heavy mass region.
The superfluid fraction $\rho_s$ in the ground state is considered to be a  predisposition that causes  the macroscopic wave nature aspect, the condensation aspect of the duality, of the $\alpha$ cluster states in the excited energy region.

\begin{figure}[t]
\begin{center}
\includegraphics[width=8.6cm]{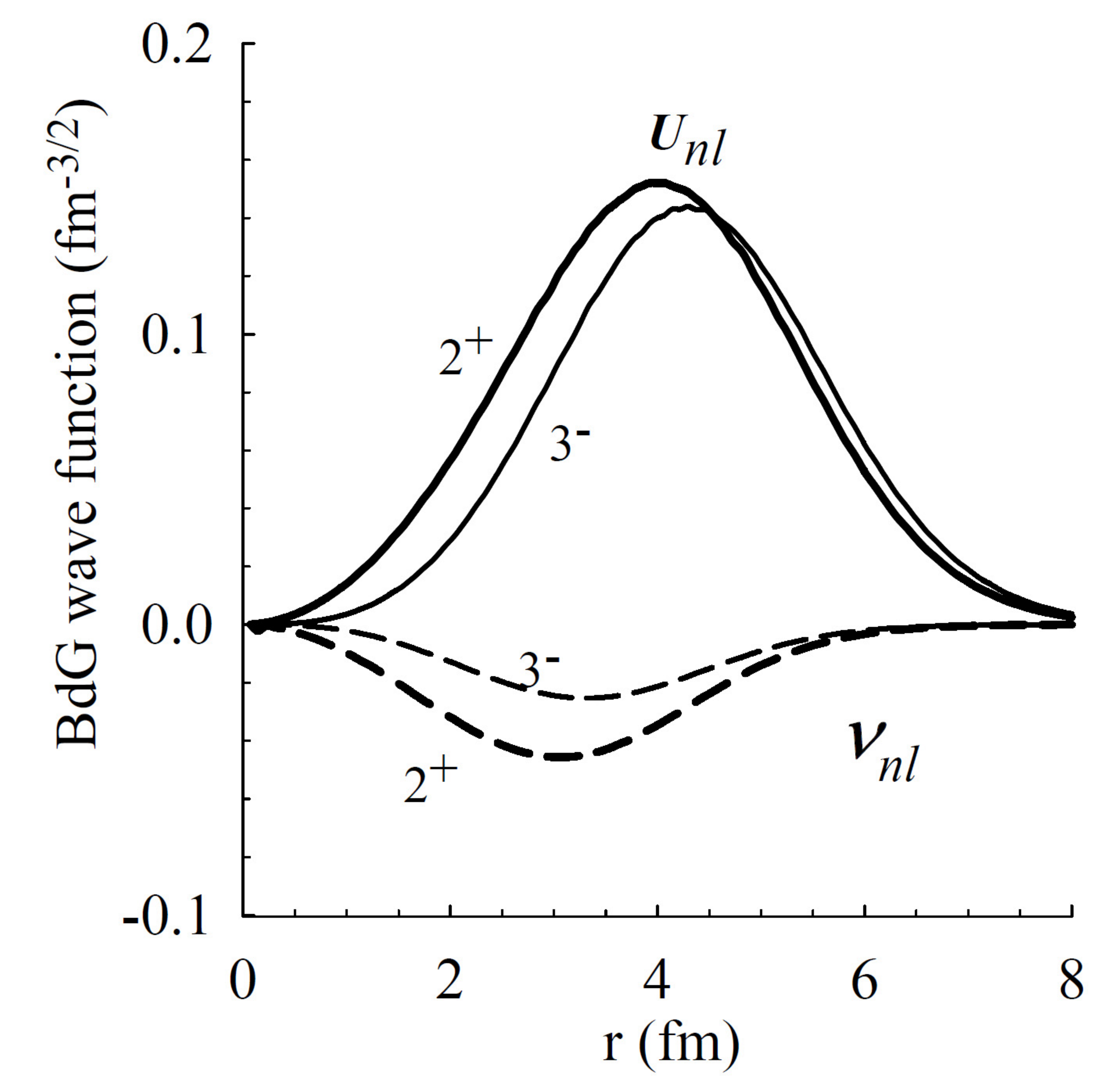}
\end{center}
\protect\caption{  Calculated  BdG wave functions of $^{40}$Ca, $\mathcal{U}_{n\ell}(r)$ (solid lines) and $\mathcal{V}_{n\ell}(r)$  (dashed lines),   for the   $2^+$ ($n=0$, $\ell$= 2) and   $3^-$  ($n=0$, $\ell$= 3) states.
}
\label{fig:BdGwf}

\end{figure}
  
  \par
  In Fig.~\ref{fig:BdGwf} the BdG wave functions $\mathcal{U}_{n\ell}(r)$ and $\mathcal{V}_{n\ell}(r)$   of Eq. (\ref{BdGsolution})   for the  $2^+$ and $3^-$ states are displayed.
The peak of $\mathcal{U}_{n\ell}(r)$
for $\ell \neq 0$  is located in the surface region because of the repulsive force between the  $\alpha$ clusters  and moves outward with increasing $\ell$ due to the centrifugal force.
The magnitude of $\mathcal{V}_{n\ell}(r)$ is  negligible  for the  $2^+$ and $3^-$ states, implying no Bogoliubov mixing  in these states  due to the  small condensation rate.

\begin{figure*}[t]
\begin{center}
\includegraphics[width=17.2cm]{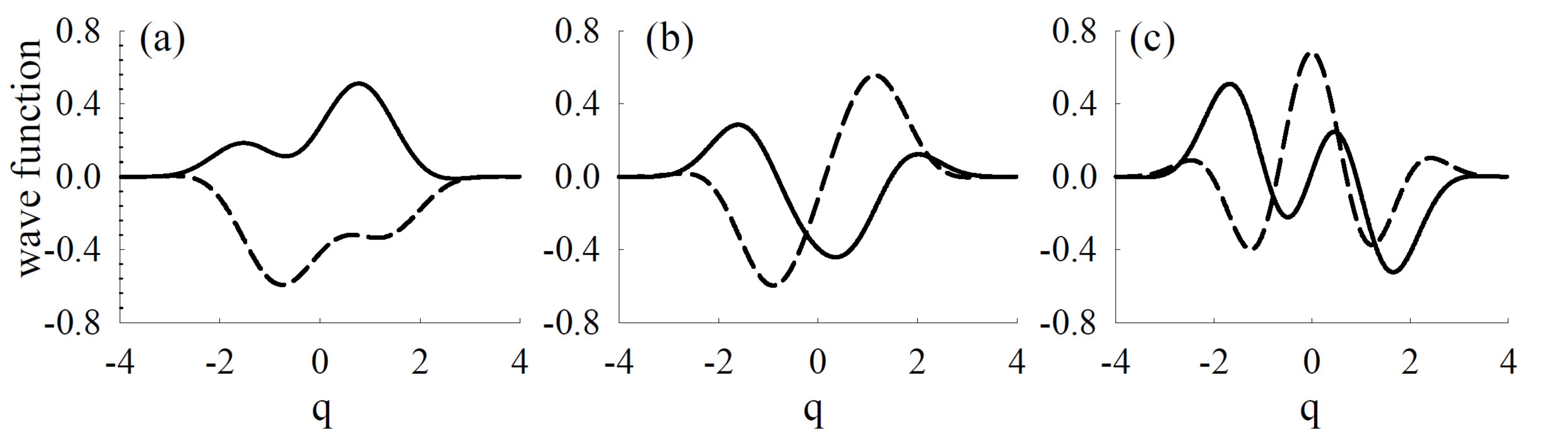}
\end{center}
\protect\caption{ The  zero-mode wave functions  $\Psi_\nu(q)$ for the $0^+$ states  of $^{40}$Ca calculated  with the condensation rate 6 \%, $N_0=0.06N$,  (a)  $\nu=0$, (b) $\nu=1$, and (c) $\nu=2$.   The solid lines and the dashed lines represent the real part and the imaginary part of  the wave functions $\Psi_\nu(q)$. 
}
\label{fig:Zeromodewf}

\end{figure*}
As for the zero-mode wave function, I introduce the eigenstate of  $\hat{Q}$, denoted by $|q>$, as $\hat{Q}$
$|q> = q|q>$. To solve Eq. (\ref{eq:HuQPeigen}), I move to the $q$-diagonal
representation, in which the state is represented by the wave
function $\Psi_\nu (q) = <q|\Psi_\nu>$, and the operators   $\hat{Q}$ and 
$\hat{P}$ are represented by $q$ and $\frac{1}{i}\frac{\partial}{\partial q}$ , respectively, consistent with the commutation relation,  [$\hat{Q}$,$\hat{P}]=i$.
In Fig.~\ref{fig:Zeromodewf} the    zero-mode wave functions $\Psi_\nu(q)$ for the first three states obtained by solving Eq. (\ref{eq:HuQPeigen}) are displayed. Figure ~\ref{fig:Zeromodewf}(a) corresponds to the ground state  with $\nu$=0 and 
Fig.~\ref{fig:Zeromodewf}(b) corresponds to the second  state with $\nu$=1, the mysterious $0^+$ state at 3.35 MeV. Figure~\ref{fig:Zeromodewf}(c)  corresponds to the third member state with $\nu$=2, $ 0^+_4$ at 7.30 MeV. 
One sees that the excitation of the NG mode is caused
by the nodal excitation of $\Psi_{\nu}(q)$ with respect to $q$ in the NG subspace. 
It is important to note that this nodal excitation is anharmonic as seen 
in $\hat H_u^{QP}$ in Eq.~(\ref{eq:HuQP}), which brings the excitation 
energy of the $\nu=1$ state  lower and  closer to the vacuum, and the $\nu=2$ state 
closer to the $\nu=1$ state in Fig.~\ref{fig:SCMEnergyLevel}. 
The  importance of the zero-mode in the BEC systems  of $\alpha$ clusters is discussed in detail  in Ref.\cite{Katsuragi2018}.

\section{DISCUSSION} 
\begin{figure}[t]
\begin{center}
\includegraphics[width=8.6cm]{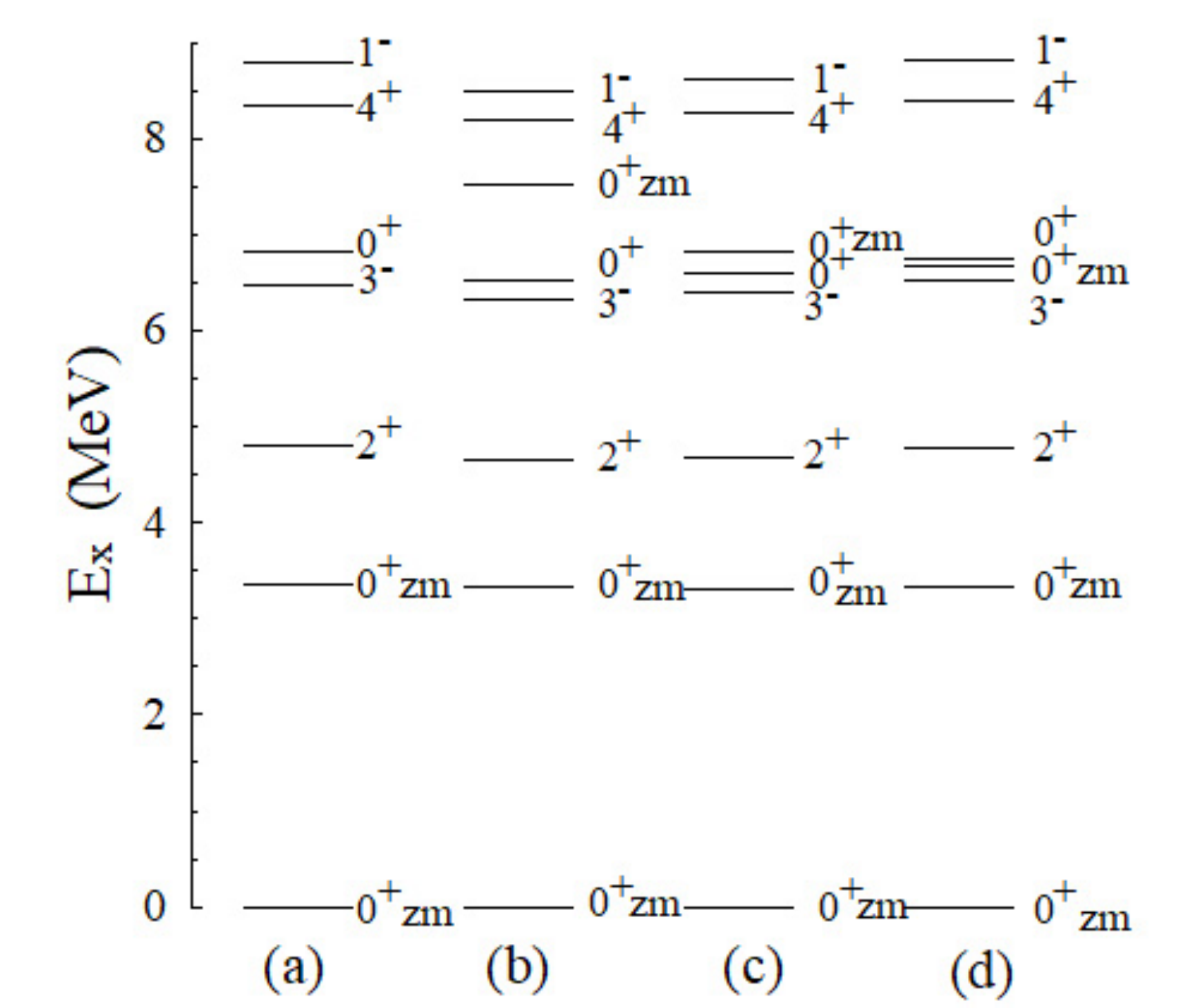}
\end{center}
\protect\caption { The condensation rate  dependence of the  energy levels  of $^{40}$Ca calculated by the superfluid   cluster model with the condensation rates  (a) 5\%,     (b) 6\%,   (c) 7\%, and   (d) 8\%. The three zero-mode states with $\nu$=0, 1, and 2 are indicated by zm and others are BdG states.
 }
\label{fig7}
\end{figure}

I study the condensation rate dependence of the calculated energy levels.
 In Fig.~\ref{fig7} the energy levels calculated for different condensation rates, 5\%, 6\%,  7\%,  and 8\%, are displayed.  The confining potential parameter $\Omega$ and the repulsive potential $V_r$ are slightly adjusted  in order to prevent the system from collapsing and the excitation energy of first excited $0^+$ state corresponds to the experimental energy, $\Omega$=3.14 MeV/$\hbar$ and $V_r$=696 MeV for 5\%,  $\Omega$=2.99 MeV/$\hbar$ and $V_r$=555 MeV for 7\%, and $\Omega$=3.04 MeV/$\hbar$ and $V_r$=535 MeV for 8\%.  
As the condensation decreases, the repulsive potential  becomes larger gradually to keep the system stable, preventing  collapse, while  the values of  $\Omega$ change little since they are related to the size of the ground state.
As seen in Fig.~\ref{fig7}, the structure of the energy spectrum changes generally little.
In  detail, the excitation energies of the $0_2^+$, $2^+$, $4^+$, $3^-$, and  $1^-$ states scarcely change for the different condensation rates. However, for the $0^+$  states,   the excitation energy of the zero-mode $\nu$=2 state decreases as the condensation rate increases gradually:  In the case of   5\% the excitation energy of the $\nu$=2 zero-mode state, 10.49 MeV, is higher than the BdG $0^+$ state.  In the case of 6\% the $\nu$=2 zero-mode state  comes down  to 7.51 MeV.  For 8\% the $\nu$=2 zero-mode $0^+$ state becomes lower than the BdG $0^+$ state.
 In the range of 6\% - 8\% the calculated  $0^+$ states correspond to the experimental energy spectrum.  
 
\begin{figure}[t!]
\includegraphics[width=8.6cm]{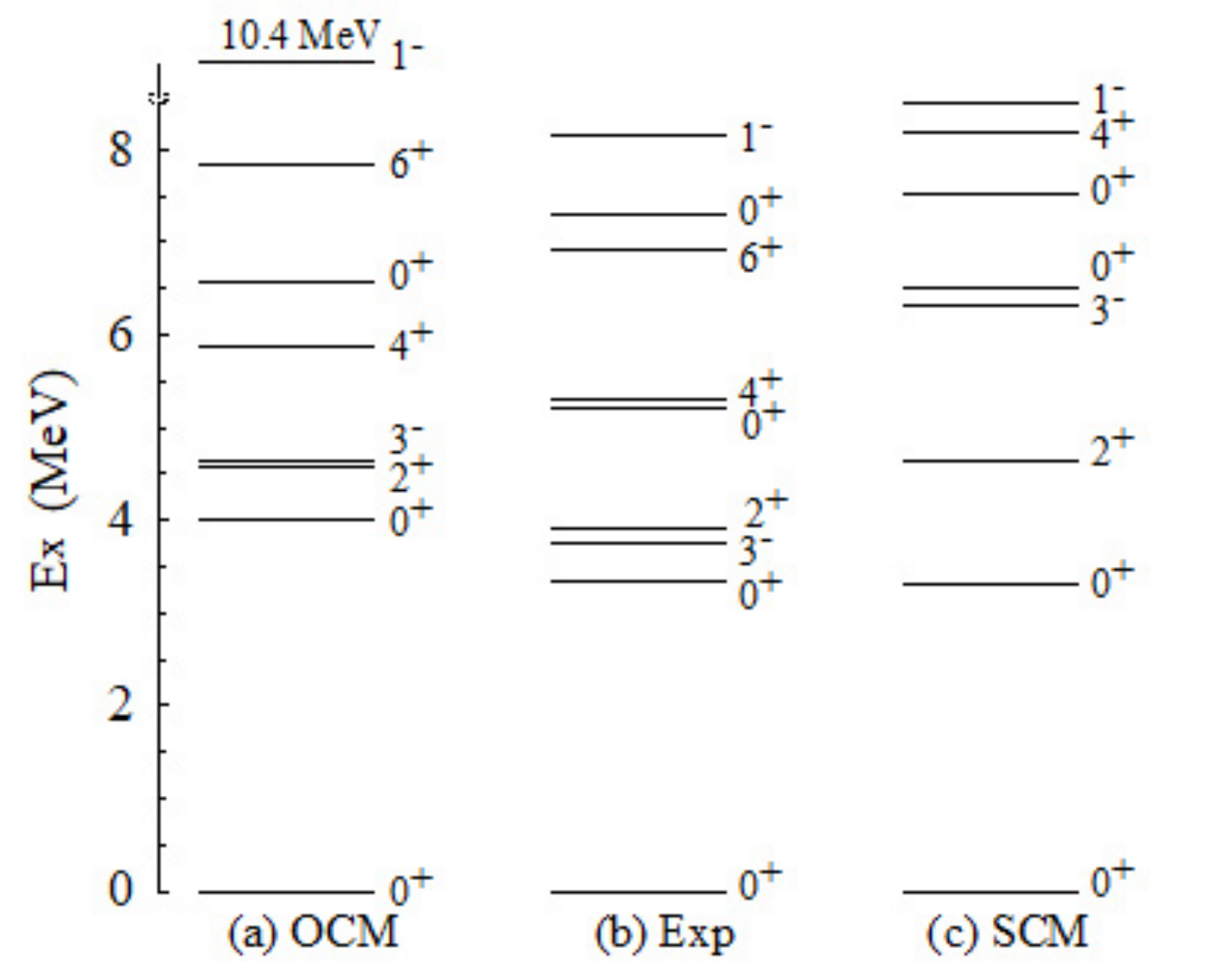}
 \protect\caption {  (a) The energy levels  calculated with the $\alpha$ cluster model with the $\alpha$+$^{36}$Ar geometrical configuration in the  OCM  \cite{Sakuda1994} are compared with (b)   the experimental energy levels \cite{ENSDF} and (c) the  field theoretical superfluid cluster model calculation with 6\% condensation rate assuming no  geometrical configuration.
 }
\label{fig8}
\end{figure}

Next I consider the crystallinity  and  the condensation aspects of the duality of the $\alpha$ cluster structure by comparing  the energy levels calculated by using  the OCM with  the $\alpha$+$^{36}$Ar cluster model in Ref. \cite{Sakuda1994} and those  by using the SCM.
 In Fig.~\ref{fig8},  both models reproduce the very low-lying mysterious $0^+$ state in agreement with experiment. 
While the OCM describes the mysterious $0^+$ state as a  deformed  state with the $\alpha$+$^{36}$Ar  geometrical configuration, which is in line with the 4p-4h dominant state in the deformed   shell model picture of   Gerace and Green \cite{Gerace1967}, the SCM describes it as a Nambu-Goldstone zero-mode state, a soft mode.  In the crystallinity picture, the mysteriously low excitation energy is brought about by the energy gain due to  deformation caused by the  geometrical $\alpha$ clustering. This mechanism  is common to  the deformed shell model  by  Gerace and Green in Ref. \cite{Gerace1967}, in which the    deformation is 
 not due to  crystallinity but 
 due to the deformation of the mean field of the shell model. It is   to be noted  that the observed significant  $\alpha$ spectroscopic factor of the mysterious $0^+$ state, $S^2_\alpha$=0.26,  is  explained by taking into account  the deformation due to  $\alpha$ clustering \cite{Sakuda1994}.
In the OCM the predisposition of $\alpha$ clustering is implemented in the ground state vacuum. In fact, the calculated ground state has the  $\alpha$ spectroscopic factors $S^2_\alpha$=0.086 for the  ($I$$\otimes$$L$)=(0$\otimes$0)
  channel  with the  [$^{36}$Ar($I$) $\otimes$ $\alpha_{L}]_{J=0}$ configuration,
 where $L$ is the orbital angular momentum of the relative motion between $^{36}$Ar and $\alpha$ clusters in $^{40}$Ca.  
 On the other hand, in the SCM the emergence of the mysterious  low excitation energy $0^+$ state is a manifestation of the emergence of a  soft mode due to  spontaneous symmetry breaking of the global phase of the ground state,  condensation aspect of the duality.

The moment of inertia of the $K=0^+$ band  calculated in the OCM is in agreement with the experimental value while that of the SCM is  slightly smaller than that in the OCM calculation. The SCM  locates the $2^+$ and $4^+$ states at  excitation energies higher than the experiment.
  However, while OCM describes the band as a rotational band of the geometrical 
    $\alpha$+$^{36}$Ar cluster structure, the SCM gives the band as the BdG  states.  Although  the two models,  two views, are completely different, they   both basically describe the $\alpha$ cluster  structure aspects of the $K=0^+$  band of $^{40}$Ca. 

As for the other excited $0^+$ states  with the $\alpha$ cluster structure above $E_x$=5 MeV, the OCM calculation locates only one  $0^+$ state between 5  and 8 MeV, which has 
  the configuration [$^{36}$Ar($2^+$) $\otimes$ $\alpha_{
L=2}]_{J=0}$.
More   $0^+$ states will be obtained in the extended 
$\alpha$+$\alpha$+$^{36}$Ar cluster model, since   the existence of a $0^+$ state with 8p-8h character has been  suggested in the deformed model and shell  model calculations in Refs. \cite{Gerace1969,Sakakura1976} and in  the $^8$Be transfer reaction experiment in Ref. \cite{Middleton1972}. 
 The SCM locates the two $\alpha$ cluster $0^+$ states in the relevant energy region; one is a zero-mode state and the other is a BdG state.

As for the negative parity states, the OCM locates  the $1^-$ state of the $K=0^-$ band with the $\alpha$+$^{36}$Ar  structure, which is   the parity-doublet partner of the $K=0^+$ band. As seen in Fig.~\ref{fig8}, its calculated excitation is slightly higher than the observed state.  On the other hand, the SCM locates the $1^-$ state as a BdG state, whose excitation energy is  in good agreement  with the experiment.
As for the  $3^-$ state, the OCM locates a low-lying $3^-$ state, which is a superposition of many channels with the   [$^{36}$Ar($I$) $\otimes$ $\alpha_{
L}]_{J=3}$ configuration with the components, 0.047, 0.055, 0.037, 0.063, 0.047, 0.036. 0.034,  and 0.071 for the channel ($I$$\otimes$$L$)=(0$\otimes$3), (2$\otimes$1), (2$\otimes$3), (2$\otimes$5), (4$\otimes$1), (4$\otimes$3), (4$\otimes$5), and (4$\otimes$7), respectively.  On the other hand, the SCM calculation locates the first $3^-$ as a BdG state at an excitation energy slightly higher than the experiment. The two results seem to be consistent that this $3^-$ state has a vibrational character.

Considering that the SCM describes the $\alpha$  clustering aspects in view of   wave nature,  the $\alpha$ cluster structure in $^{40}$Ca is found to be understood from  both the viewpoints of crystallinity and condensation associated with superfluidity, a property of supersolidity.
It is found in the SCM that  the emergence of the $0^+$ state at  very low  excitation energy, which is mysterious   from the viewpoint of multiparticle multihole excitation in the shell model, is the consequence of the first excited $0^+$ being a  collective state with a  soft mode nature caused by the  NG zero mode due to the spontaneous symmetry breaking of the vacuum ground state due to the  condensation aspect,   superfluidity, of the duality of  $\alpha$ clustering. 
 This mechanism is quite general when  SSB of the global phase of the vacuum occurs. In the previous paper \cite{Ohkubo2020} the mysterious $0^+$ state in $^{12}$C, the Hoyle, state was understood as a  soft mode state of the zero mode  due  SSB     due to the condensation aspect of the duality of $\alpha$  cluster structure of the vacuum ground state. 

What is the evidence for the  supersolidity? 
The direct evidence of  supersolidity is the observation of 
  a Nambu-Goldstone mode \cite{Nambu1960,Goldstone1960,Nambu1961} due to SSB of the global phase as   was confirmed  very recently for an optical lattice supersolid \cite{Tanzi2019A,Natale2019,Guo2019}. 
  Since the superfluid density  $\rho_s$ is the order parameter of the SSB of the global phase  \cite{Ohkubo2020}, the existence of $\rho_s$$\ne0$ in the GCM $\alpha$ cluster wave function of the ground state due to the duality  accompanies  a Nambu-Goldstone mode state, which is  a very low-lying collective state and is difficult to explain in the shell model.  
   This logic is  same as  the emergence of  rotational band states in deformed nuclei.    
Quadrupole  deformation with the  order parameter $\delta\ne0$ is   caused  by  a quadrupole boson condensation in  the ground state due to SSB of  rotational invariance  \cite{Ring1980}.
 The  appearance of the    intruder collective states at a very  low excitation energy near the $\alpha$ threshold such as the  mysterious  $0_2^+$      states in  $^{40}$Ca     and in  $^{16}$O, analogous to the intruder $0_2^+$ state in $^{12}$C, which has been understood by the empirical threshold rule of  the Ikeda diagram \cite{Ikeda1968,Horiuchi1972}, is considered to be  understood from the  viewpoint of  the Nambu-Goldstone mode due to SSB of the global phase of the $\alpha$ cluster structure.

It is to be noted that the present reasonable success of the SCM, which assumes no geometrical $\alpha$ cluster configuration, does not mean  ruling out the geometrical $\alpha$ cluster model. It should be emphasized that the SCM only describes the condensation, superfluid, aspect of the duality of the $\alpha$ cluster structure, being complementary to the geometrical $\alpha$ cluster picture. In the $\alpha$ cluster structure the geometrical structure is essential. The SCM does not replace the geometrical $\alpha$ cluster model. For example, the rotation motion, the large moment of inertia,  is caused by the spontaneous symmetry breaking of the rotational invariance  due to the  geometrical $\alpha$ cluster configuration, which is absent in the present SCM under a spherical trapping potential.
The reduction of the moment  of inertia in the SCM calculations is inherent to superfluidity, which is  well known in the superfluid heavy nuclei \cite{Ring1980}. 
A cluster model with a geometrical configuration which involves the order parameter to characterize the condensation of $\alpha$ clusters is a future work to be studied.


Finally I mention  the importance of the Pauli principle for the duality of geometrical crystallinity  and  condensation  of $\alpha$ cluster structure.  The geometrical crystallinity of the  $\alpha$ clusters has been known to be  caused by the Pauli principle \cite{Tamagaki1969,Ohkubo2016}. 
In  Fig.~1(c), 1(f), 1(i), and 1(l) of each nucleus the coherent wave    of the  $\alpha$ cluster structure   is the consequence of  the  geometrical  crystallinity.
Thus the Pauli principle   has  the dual role of causing  the 
  geometrical clustering and condensation.
 In this sense the origin of the superfluidity of  $\alpha$ cluster structure is  different from that of  the BCS superfluidity in heavy nuclei and cold atoms.

 \section{Summary} 
 \par
It was shown that   the spatially localized Brink $\alpha$ cluster model  in the  generator coordinate method has the apparently incompatible duality of  crystallinity and      condensation of $\alpha$ clusters,  a property of  a supersolid. 
In order to see whether the $\alpha$ cluster structure, which in recent decades has been understood based on the crystallinity picture with a geometrical configuration of clusters, is also understood from the other aspect of the duality, condensation, a field theoretical  cluster  model is used in which the order parameter of condensation is introduced.
  The $\alpha$ cluster structure of $^{40}$Ca with  a mysterious $0^+$ state  at   very  low excitation energy was investigated by the SCM with ten $\alpha$ clusters. 
 Since the SCM rigorously treats  spontaneous symmetry breaking of the global phase, a Nambu-Goldstone collective mode, zero mode,  due to condensation inevitable appears.
It is shown that  the mysterious $0^+$ state, which is considered to be a band head state of the $K=0^+$ band with the $\alpha$+$^{36}$Ar cluster structure in the crystallinity picture, is understood   as an NG zero-mode  state. 
The $1^-$ state, which is considered in the crystallinity picture to be a band head state of the $K=0^-$ band, which is a  parity-doublet partner of the $K=0^+$ band, is obtained as a BdG state in correspondence to the experimental data.
The two $\alpha$ cluster $0^+$ states at around 6 MeV are also obtained in accordance with the experimental data. 
Thus it is found that the low-lying $\alpha$ cluster states, which have been considered to be understood in the geometrical cluster picture, can be also understood from the other aspect of the duality, condensation.
The mysterious $0^+$ state of $^{40}$Ca is a collective mode, a soft mode, of the  
NG mode  caused by SSB of the ground state vacuum. 
This explains  naturally why the   mysterious low-lying $0^+$ state appears below or near the $\alpha$ threshold energy of $^{40}$Ca.
  This mechanism is  logically the  same as  the emergence of the NG mode  rotational band states in deformed nuclei, which is 
 caused by  a quadrupole boson condensation
 due to SSB of  rotational invariance  \cite{Ring1980}.
 The  appearance of  such   intruder collective states   near the $\alpha$ threshold, which has been understood by the empirical threshold rule of  the Ikeda diagram \cite{Ikeda1968,Horiuchi1972}, is understood to be due to the NG   mode state due to condensation  of the $\alpha$ cluster structure.
  The dual property of   crystallinity and condensation,  a property of  a supersolid, of  $\alpha$ cluster structure may be a general  feature of the $\alpha$ cluster structure.
  Since the Pauli principle is responsible  for clustering \cite{Ohkubo2016,Tamagaki1969},  one can say that  supersolidity of $\alpha$ cluster structure  is the consequence of  the  Pauli principle.

The author  thanks J. Takahashi for the numerical calculations using the superfluid cluster model in the early stage of this work and Y. Yamanaka for interest. He also thanks  the Yukawa Institute for Theoretical Physics, Kyoto University for   the hospitality extended  during  a stay in  2019.

\end{document}